\documentclass[pre,reprint,amsmath,amssymb,showpacs]{revtex4-1}
\usepackage{dcolumn}
\usepackage{bm}
\usepackage[dvips]{color}
\usepackage{subfigure}
\usepackage{graphicx} 
\usepackage{algorithmicx}
\usepackage{pict2e}
\usepackage[ruled]{algorithm}
\usepackage{algpseudocode}
\newcommand{\rvec}{\textit{\textbf{r}} }

\begin{document}

\title{Rigidity of transmembrane proteins determines their cluster shape}

\author{Hamidreza Jafarinia$^{1}$}
\author{Atefeh Khoshnood$^2$}
\author{Mir Abbas Jalali$^{3}$}
\email{mjalali@berkeley.edu}
\affiliation{
$^1$Department of Mechanical Engineering, Sharif University of Technology, P.O. Box: 11155-9567, Tehran, Iran \\
$^2$Reservoir Engineering Research Institute, Palo Alto, California 94301, USA \\
$^3$Department of Astronomy, University of California, Berkeley, California 94720, USA}


\begin{abstract}
Protein aggregation in cell membrane is vital for the majority of biological functions. Recent experimental results suggest that transmembrane domains of proteins such as $\alpha$-helices and $\beta$-sheets have different structural rigidities. We use molecular dynamics simulation of a coarse-grained model of protein-embedded lipid membranes to investigate the mechanisms of protein clustering. For a variety of protein concentrations, our simulations under thermal equilibrium conditions reveal that the structural rigidity of transmembrane domains dramatically affects interactions and changes the shape of the cluster. We have observed stable large aggregates even in the absence of hydrophobic mismatch, which has been previously proposed as the mechanism of protein aggregation. According to our results, semi-flexible proteins aggregate to form two-dimensional clusters, while rigid proteins, by contrast, form one-dimensional string-like structures. By assuming two probable scenarios for the formation of a two-dimensional triangular structure, we calculate the lipid density around protein clusters and find that the difference in lipid distribution around rigid and semiflexible proteins determines the one- or two-dimensional nature of aggregates. It is found that lipids move faster around semiflexible proteins than rigid ones. The aggregation mechanism suggested in this paper can be tested by current state-of-the-art experimental facilities.
\end{abstract}


\maketitle


\section{Introduction}
\label{sec:intro}

Transmembrane (TM) proteins are regulators of several cellular processes. In order to perform their functions, they often aggregate and distribute non-uniformly in the cell membrane \cite{sieber2007anatomy}. The aggregation process of proteins sometimes becomes abnormal, and causes amyloid diseases \cite{chiti2006protein}. In recent years, there has been increasing interest in identifying various mechanisms that affect protein--membrane interactions \cite{mouritsen1984mattress, west2009membrane} and, consequently, the aggregation behavior of different proteins.

The specific structure of TM proteins defines their physical properties and enables them to perform their biological functions \cite{buehler2006nature}. TM proteins differ in size and physical properties \cite{becker2009biogenesis}, and may have single or multiple $\alpha$-helical \cite{bocharov2010structure, westfield2011structural} or $\beta$-structure \cite {tang2013desalination} domains. Recent experimental studies show that $\alpha$-helical structures have 
softer domains than $\beta$-structures in both dry and hydrated states, and $\beta$-barrels and $\beta$-sheets are more rigid structural units than $\alpha$-helices \cite{perticaroli2014rigidity}. The higher number of hydrogen bonds per residue is also considered to be the probable cause of the more rigid structure of $\beta$-sheets compared to $\alpha$-helices \cite{perticaroli2013secondary}. It has also been shown that the secondary structure of proteins affects the rigidity and dynamics of the protein. Proteins containing $\beta$-structures have a higher Young's modulus and higher frequency of the collective vibration \cite {perticaroli2013secondary}. Consequently, the effect of the class and structural rigidity of proteins on their aggregation behavior and biological function cannot be ignored.   

Although lipid raft formation \cite{mcintosh2006roles} and direct linking \cite{feng1998dual} of proteins cause the aggregation of membrane proteins, membrane curvature \cite{bahrami2011vesicle} and membrane-mediated interactions play important roles on the formation and fragmentation of protein clusters. Among mechanisms that generate lipid-mediated protein interactions, the hydrophobic mismatch interaction, which is due to the difference between the hydrophobic lengths of the integral proteins and the hydrophobic thickness of their host membrane, has been widely studied by several groups \cite{venturoli2005simulation,schmidt2008cluster,de2008molecular,west2009membrane}. However, we know little about the effect of the structural properties of proteins on the cluster formation. It is known that the shapes and sizes of proteins determine the distribution of their surrounding lipid molecules \cite{de2008molecular, schmidt2010hydrophobic,morozova2012shape, yoo2013membrane}, which in turn affect the stability and function of proteins \cite{jensen2004lipids, lee2004lipids}. What is poorly understood is how the structural rigidity of proteins integrates with other factors to shape the patterns of aggregates. 

In this paper, we use coarse-grained molecular dynamics simulations to systematically investigate the effect of structural rigidity of TM proteins on the formation of clusters. We design specific model proteins to exclude the effect of hydrophobic mismatch and isolate the role of structural rigidity. We use two sets of proteins: semiflexible and rigid. In \S\ref{sec:model}, we present our model and simulation method and setup. In \S\ref{sec:results}, the results of molecular dynamics simulations are presented. We discuss our findings in \S\ref{sec:discuss} and show how they are comparable with experimental observations and previous theoretical modelings. Our simulations show that proteins form clusters even in the absence of hydrophobic mismatch and the final shapes of protein aggregates in lipid bilayers depend strongly on the rigidity of proteins.
 
\section{Model and Methods}
\label{sec:model}

\begin{figure}
\centerline{\hbox{\includegraphics[width=0.4\textwidth]{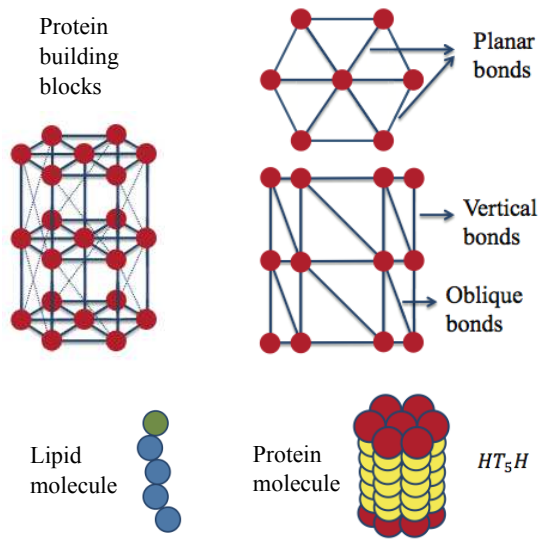}}}
\caption{(Color online) Models of protein and lipid molecules and their bonds. In the three-dimensional perspective view of the protein, oblique bonds are shown by dashed lines. Hydrophilic and the hydrophobic particles of the $HT_{5}H$ protein are depicted as red and yellow spheres, respectively. Hydrophilic and hydrophobic particles of lipid molecules are represented by green and blue spheres, respectively. These coloring schemes are also used in the snapshots of simulations.}
\label{fig1}
\end{figure}

The model of lipid molecules is composed of one hydrophilic head particle and a hydrophobic tail chain \cite{goetz1998computer, reynwar2007aggregation}, which contains four particles. Our TM proteins are modeled as hexagonal prisms with middle hydrophobic particles and hydrophilic groups at both ends \cite{schmidt2008cluster, schmidt2010hydrophobic}. The hydrophobic mismatch is tuned by changing the length $\Delta r=r_p-r_l$ of the hydrophobic part of proteins, where $r_{p}$ is the length of the hydrophobic part of proteins, and $r_{l}$ is the average thickness of the hydrophobic part of the bilayer. Models of lipid and protein molecules and their corresponding bonds are displayed in Fig. \ref{fig1}. Hydrophilic and hydrophobic particles are labeled H and T, respectively. In this study, our length scale is $\sigma=1/3$ nm and the energy unit is $N_{\rm A}\epsilon=2$ kJ/mol, with $N_{\rm A}$ being the Avogadro number. In each lipid or protein molecule, the adjacent $i$th and $(i+1)$th particles interact through the harmonic bond potential
\begin{equation}\label{six}
U_b(r_{i,i+1})=k_b(r_{i,i+1}-\sigma_{eq})^2,
\end{equation}
where $r_{i,i+1}$ is the distance between particles and $\sigma_{eq}$ is the equilibrium length of the bonds. Three kinds bonds--- planar, vertical, and oblique---are used to build two types of protein molecules---rigid and semiflexible. The planar and vertical bonds have identical spring constants of $k_{b1}=5000 \, \epsilon/\sigma^{2}$, which is fixed in all simulations. For the oblique bonds of rigid proteins (RPs) and semiflexible proteins (SFPs) we set $k_{b1}=5000 \, \epsilon/\sigma^{2}$ and $k_{b2}=35 \, \epsilon/\sigma^{2}$, respectively. Both SFPs and RPs are more rigid than lipid molecules. We set $\sigma_{eq}=\sigma$ for lipid bonds and planar and vertical bonds of proteins. For oblique protein bonds, we use $\sigma_{eq}=\sqrt{2} \, \sigma$. SFPs are stiffer than lipid molecules and mostly maintain their hexagonal cross section when they are bent due to interactions with other proteins and lipids. The angle between consecutive bonds in a lipid molecule is controlled by
\begin{equation}
U=k_{a}(\cos\theta-cos\theta_{eq})^2,
\end{equation}
where $k_{a}=1.85\,\epsilon$ and $\theta_{eq}=\pi$.
Interactions between the particles of different molecules are governed by soft-core and Lennard-Jones potentials,
defined as
\begin{eqnarray}
U_{\rm sc}(r_{ij}) &=& 4\epsilon \left (\frac{\sigma_{sc}}{r_{ij}} \right )^9, \label{eq1} \\
U_{\rm LJ}(r_{ij}) &=& 4\epsilon \left [\left (\frac{\sigma}{r_{ij}} \right )^{12}-\left ( \frac{\sigma}{r_{ij}} \right )^6 \right ], \label{eq2}
\end{eqnarray}
where $r_{ij}=\vert \rvec_i-\rvec_j \vert$ and $\sigma_{\rm sc}=1.05 \, \sigma$. In these equations, $\rvec_i$ is the global position vector of the $i$th particle. The repulsive soft-core potential is used to model the interaction between hydrophobic and hydrophilic particles, and between solvent and hydrophobic particles. All other interactions are modeled by the Lennard-Jones potential. A cutoff radius of $r_c=2.5 \, \sigma$ is applied to $U_{\rm sc}$ and $U_{\rm LJ}$, which are then shifted in order to vanish at $r_{ij}=2.5 \, \sigma$ \cite{goetz1998computer}. This guarantees the continuity of both potential fields. 

We carry out molecular dynamics simulations of $NVT$ ensembles using the \textsf{\small LAMMPS} package. Periodic boundary conditions are imposed and the temperature is kept constant at $T_0=310$ K (with $k_B T=1.29 \, \epsilon$) utilizing the Nose-Hoover thermostat. Here $k_B$ is the Boltzmann constant and the lipid bilayer is in the liquid phase. All particles have the same mass, $N_{\rm A}m=36$ gr/mol \cite{goetz1998computer}. The integration time step is set to $\Delta t=0.005 \, \tau$, where $\tau=\sqrt {m\sigma^2/\epsilon}$ is the intrinsic time scale. The actual value of $\Delta t$ is $7.07$ fs. In all simulations, the number density of particles is $n=0.66/\sigma^{3}$ and the area per lipid (for a bilayer without proteins) is $A_s=2.09 \, \sigma^{2}$. $A_s$ is the area of lipid bilayer divided by the number of lipids. We select the number of lipids such that bilayers with minimum surface tension and without permanent curvatures are obtained. This helps us to eradicate the effect of curvature-mediated interactions. Under this condition the surface tension of the bilayer is positive and approximately equals $0.24\, \epsilon\sigma^{-2}$, corresponding to 7 mN/m. This has led to stable bilayers in all of our simulations. It is remarked that the rupture surface tension of biological membranes varies from 1 to 30 mN/m, and it depends on the lipid composition of the bilayer \cite{evans2003dynamic,needham1990elastic}. Simulations are performed for various concentrations of proteins embedded in the bilayer. We denote the concentration of proteins $c_p=N_{p}/(N_{p}+N_{l})$, where $N_{p}$ and $N_{l}$ are the numbers of protein and lipid particles, respectively. At the beginning of simulations, each protein molecule is placed in the bilayer by removing nine lipid molecules. 

\section{Results}
\label{sec:results}

We investigate the clustering process for low and high protein concentrations and for two protein types: RPs and SFPs. To quantify the flexibilities of proteins, we have compared the longitudinal flexibilities of our protein models with each other and, also, with a lipid patch that occupies the same area that proteins do. The torsional flexibilities of protein models have also been computed. 

We have carried out simulations for a single RP and SFP in vacuum as well as in a lipid bilayer and calculated the standard deviation of the length of the hydrophobic part for each protein. A bilayer consisting of nine lipid molecules in a box of size $3.06 \sigma \times 3.06 \sigma \times 10\sigma$ was simulated. The area of this bilayer patch approximately equals the area of proteins, and it has the same area per lipid ($A_s=2.09 \sigma^2$) as other bilayers in our simulations. Let us define the lengths of the hydrophobic parts of the RP and SFP by $ r_{\rm RP}$ and $r_{\rm SFP}$, respectively. Denoting the standard deviation $SD(\cdot)$, we find
\begin{eqnarray}
SD(r_{\rm RP})= 0.025\sigma,~~SD(r_{\rm SFP})=0.067\sigma, 
\end{eqnarray}
for proteins in vacuum,
\begin{eqnarray}
SD(r_{\rm RP})= 0.025\sigma,~~SD(r_{\rm SFP})=0.055 \sigma,
\end{eqnarray}
for proteins in the bilayer, and $SD(r_l)=0.32 \sigma$ for the lipid bilayer. It is seen that transmembrane RPs are stiffer than SFPs, and SFPs are stiffer than a bilayer that has the same surface area of proteins. Due to the lack of interactions with lipid and solvent molecules, SFPs are more flexible in vacuum. 

The weaker oblique bonds result in more torsional flexibility for SFPs. The torsional rigidity of proteins can be measured by the relative twist angle of their two hydrophilic ends. Defining $\theta_{\rm RP}$ and $\theta_{\rm SFP}$ as the relative twist angles of the upper and lower hydrophobic parts, we obtain
\begin{eqnarray}
SD(\theta_{\rm RP})= 0.11^{\circ},~~SD(\theta_{\rm SFP})=0.28^{\circ},
\end{eqnarray}
for proteins in vacuum and 
\begin{eqnarray}
SD(\theta_{\rm RP})= 0.11^{\circ},~~SD(\theta_{\rm SFP})=0.22^{\circ}, \end{eqnarray}
for proteins in the bilayer. These results show that RPs are torsionally stiffer than SFPs. Our findings are consistent with the higher conformational displacements of $\alpha$-helices compared to $\beta$-sheets \cite{gaspar2008dynamics,perticaroli2013secondary}.

\begin{figure}
\centerline{\hbox{\includegraphics[width=0.46\textwidth]{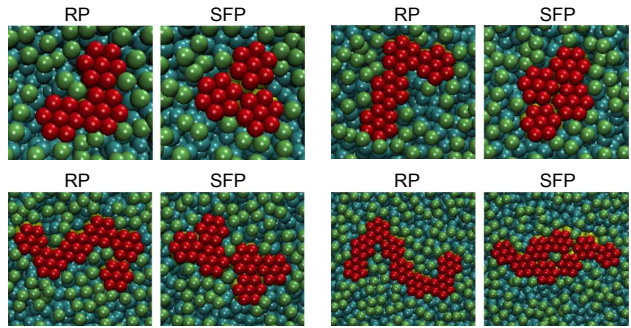}}  
                 } 
\caption{(Color online) Shapes of clusters formed by SFPs and RPs. \textit{Top left}: $N=3$. \textit{Top right}: $N=4$. \textit{Bottom left}: $N=6$. \textit{Bottom right}: $N=8$. It is evident that RPs form string-shaped clusters.}
\label{fig2}
\end{figure}

In all of our simulations, the effect of hydrophobic mismatch is neutralized, $\Delta r=0.01$, by using the $HT_{5}H$ model proteins. Figure \ref{fig2} shows snapshots of protein-embedded membranes with SFPs and RPs. We have used $N=3$, 4, 6, and 8 protein molecules for both protein types. The size of the lipid bilayer is $16\,\sigma \times 16\,\sigma$ for $N=3$, $23\,\sigma \times 23\,\sigma$ for $N=4$ and 6, and $35\,\sigma \times 35\,\sigma$ for $N=8$. The snapshots have been taken at $t=5 \times 10^{6}\Delta t$. The difference between SFP and RP clusters is prominent: RPs form string-like one-dimensional structures, while SFPs form two-dimensional clusters. These results differ significantly from previous work \cite{venturoli2005simulation}, which suggests that there is only weak attraction between inclusions in the absence of mismatch. Our results, clearly, show that protein inclusions form stable clusters in the absence of hydrophobic mismatch. One of the distinct features of SFPs is their clustering in triangular structures. For the model with six and eight SFPs, two triangular structures can be seen in Fig. \ref{fig2}. RPs do not share this property. 
 
\begin{figure*}
\centerline{\hbox{\includegraphics[trim={0.0cm 0 0 0},clip,width=0.3\textwidth]{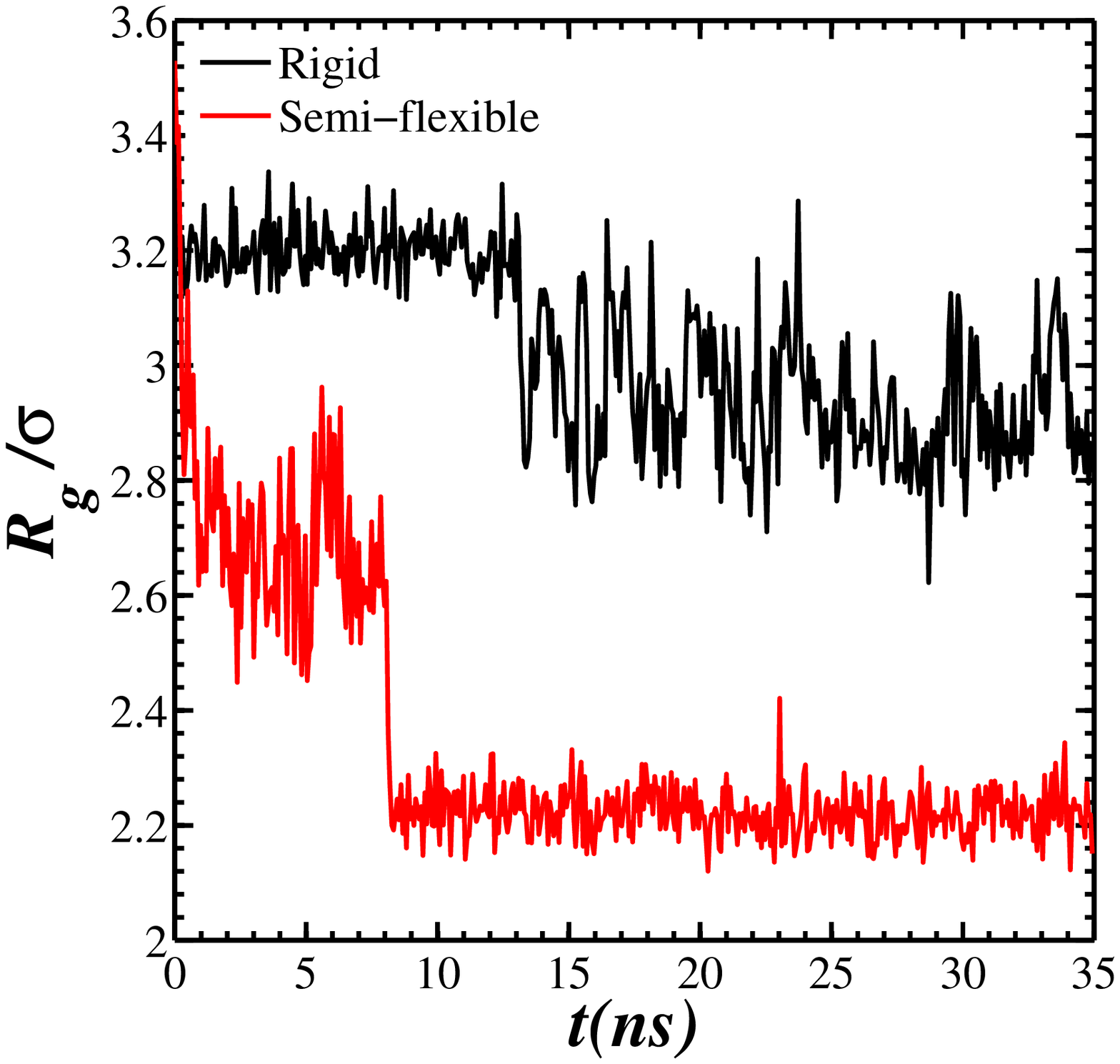}} \hspace{0.1in}
  \hbox{\includegraphics[trim={0.0cm 0 0 0},clip,width=0.3\textwidth]{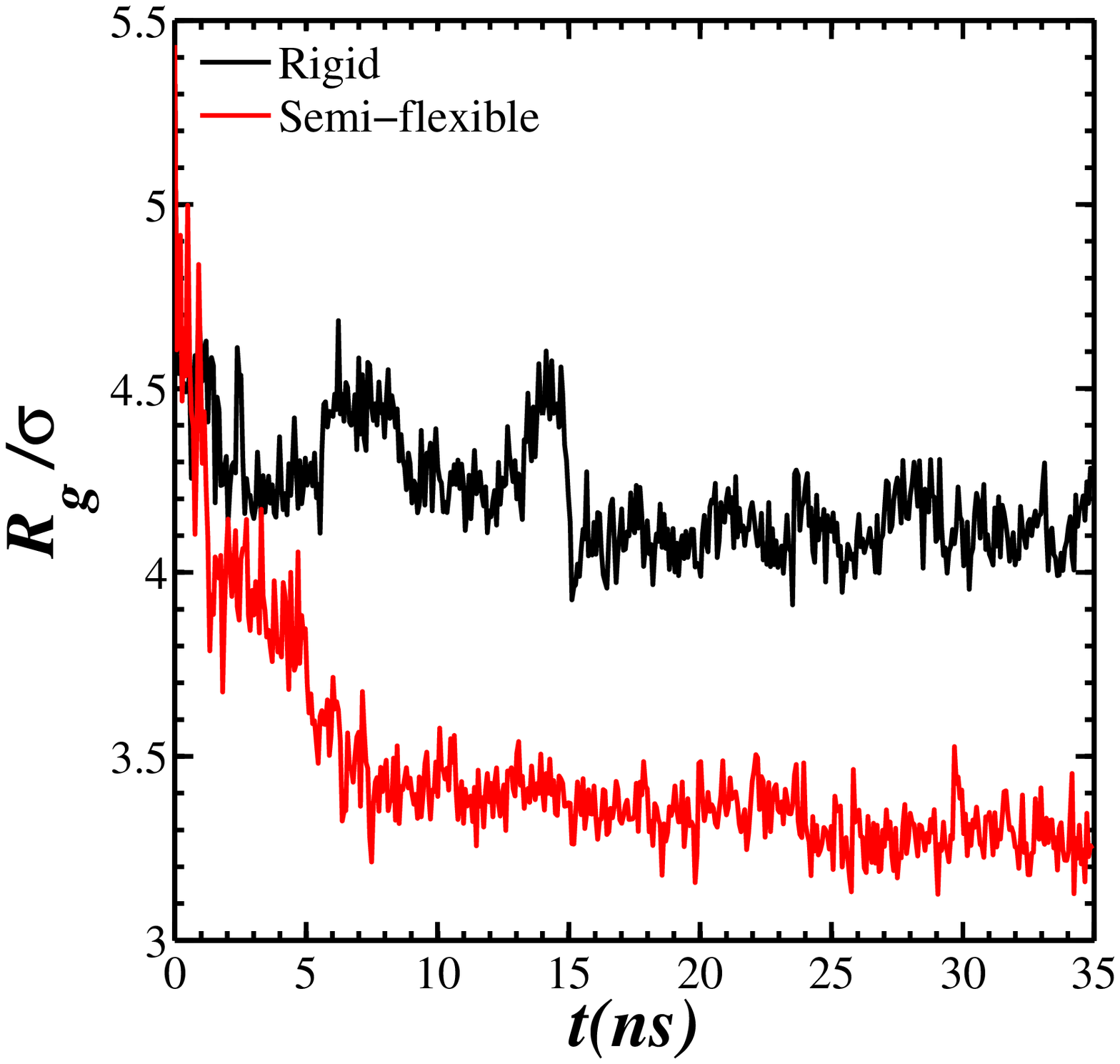}} \hspace{0.1in}
  \hbox{\includegraphics[trim={0.0cm 0 0 0},clip,width=0.3\textwidth]{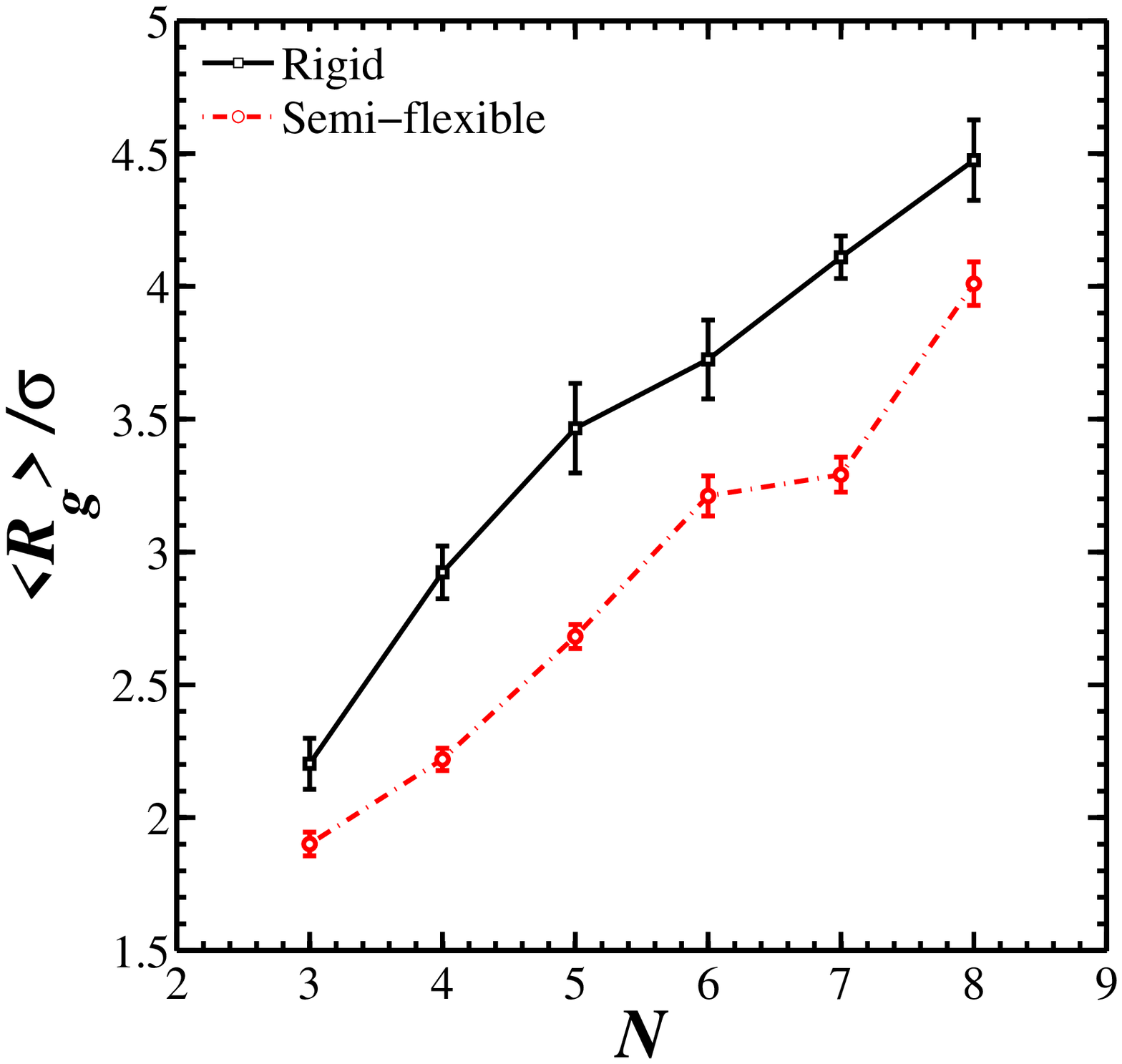}} }
\caption{(Color online) Temporal evolution of the gyration radius of protein clusters for $N=4$ (left), and $N=7$ (middle). Black lines correspond to RPs; red lines to SFPs. \textit{Right panel}: Time-averaged gyration radius versus number of proteins, $N$, in a cluster. The radius of gyration is consistently larger for RPs.}
\label{fig3}
\end{figure*}

To quantify the discrepancies between one-dimensional and two-dimensional clusters, we measure the radius of gyration of protein structures as 
\begin{equation}
R_g^2=\frac{1}{M}\sum_{i=1}^{N_{p}} m_i \left | \rvec_i-\rvec_{\rm c} \right | ^2,
\label{Rg}
\end{equation}
where $N_{p}$ and $M$ are the total number and the total mass of the head particles of proteins, respectively, and $r_{\rm c}$ and $\rvec_i$ define the position vectors of the $i$th particle and the center of mass of protein heads, respectively. The mass of each particle is denoted $m_{i}$. To compute $R_g$, we have used protein heads lying in one monolayer. The variation of $R_g$ over time has been plotted in Fig. \ref{fig3} for a system with four and seven proteins. We have also plotted the time-averaged radius of gyration, $\langle R_g \rangle =\frac 1{t_2-t_1} \int_{t_1}^{t_2} R_g(t) \, dt$, for several number of proteins in a cluster. We have used $(t_1,t_2)=(30,35\,{\rm ns})$ for both RPs and SFPs. It is shown that for the same number of proteins, two-dimensional structures formed by SFPs always have a lower radius of gyration; the radius of gyration increases proportionally to the number of proteins in the cluster. Since the linear aggregates have more contact and interaction with surrounding lipids, these structures are more likely to change their shape slightly, for example, from a completely straight linear structure to a curved line, which, in turn, alters the radius of gyration. 

As Fig. \ref{fig2} shows, string-like clusters have a variety of configurations with large variations in their gyration radii (see Fig. \ref{fig3}). In simulations of SFPs we have observed deformed cross sections of proteins (deviations from hexagonal shapes), especially when they do not belong to a cluster. Deformations of proteins can occasionally stabilize them in the membrane, so that they do not participate in cluster forming processes. So SFPs are stabilized by either contributing to a cluster or undergoing deformation while they are singly dispersed in the membrane. It is remarked that we have repeated our simulations starting with different initial conditions and observed similar clustering trends for each protein class. Moreover, we continued our simulations over longer time scales, up to 200 ns, and obtained the same results as in our 35-ns-long simulations for both SFP and RP clusters. 

\begin{figure}[b]
\centerline{\hbox{\includegraphics[width=0.48\textwidth]{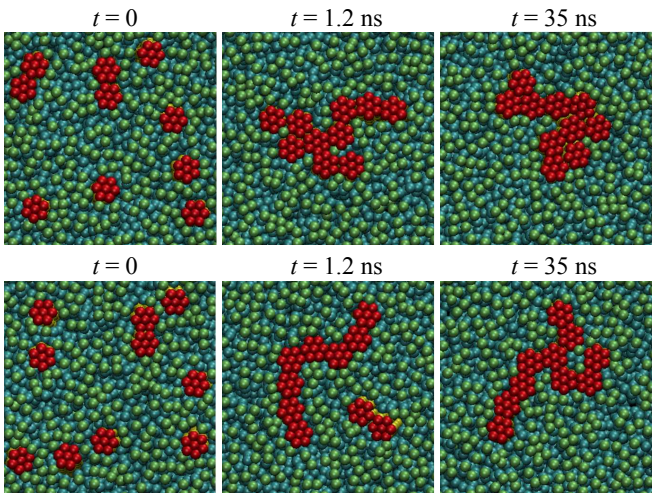}} 
}
\caption{(Color online) Clustering process for SFPs (top) and RPs (bottom). 
Time increases from left to right.}
\label{fig4}
\end{figure}

Figure \ref{fig4} shows the temporal evolution of cluster formation from an initially random distribution of proteins. Interestingly, RPs immediately aggregate to one-dimensional clusters, whereas SFPs first make a cluster with several branches separated by lipid molecules, then evolve to a compact cluster as trapped lipids are released.

\begin{figure}
\centerline{\hbox{\includegraphics[width=0.4\textwidth]{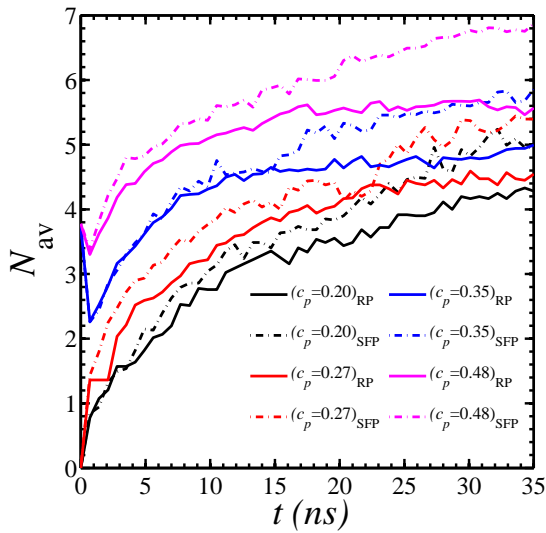}} }                  
\caption{(Color online) Evolution of the average number of neighboring proteins, $N_{\rm av}$, for SFPs and RPs and four choices of protein concentrations. The difference $\Delta N_{\rm av}=N_{\rm av,SFP}-N_{\rm av,RP}$ between SFPs and RPs increases over time as more SFPs participate in the cluster formation process.}
\label{fig5}
\end{figure}

To eliminate possible boundary effects in simulation boxes with periodic boundary conditions, and also understand how the concentration of proteins affect the aggregation process, we have carried out simulations in a larger membrane, of size $90 \sigma \times 90 \sigma$ with higher protein concentrations, $c_p=0.2$, 0.27, 0.35, and 0.48. We have studied several snapshots of these simulations and our previous conclusions for smaller membranes are unaltered. A few more results for high-$c_p$ simulations are as follows: (i) SFPs rarely form one-dimensional clusters, though the lengths of such clusters are much shorter than the aggregates of RPs; (ii) one-dimensional structures of RPs may connect to each other to form longer or branched web-like structures; and (iii) on rare occasions RPs cluster as two-dimensional domains. Observations i and iii might be due to the longer relaxation time scales of large membranes with a high $c_p$. The structure formation process by RPs and SFPs is better understood by computing the average number of neighboring proteins, $N_{\rm av}$. For a given (subject) protein $i$, we find the number of neighboring proteins $N_{i}$ within a distance of $6.1\, \sigma$, measured from the center of mass of the protein. We then calculate the average number $N_{\rm av}=\frac 1N \sum_{i=1}^{N} N_i$. Figure \ref{fig5} shows the variation of $N_{\rm av}$ over time for two protein types and several protein concentrations. It is shown that $N_{\rm av,SFP}$ is consistently larger than $N_{\rm av,RP}$, and the difference $\Delta N_{\rm av}=N_{\rm av,SFP}-N_{\rm av,RP}$ increases versus time as the aggregate size and structure reach steady state. We note that the time for $N_{\rm av}$ to reach the steady-state value is shorter in systems with RPs and at higher protein concentrations.

\begin{figure*}
\centerline{\hbox{\includegraphics[width=0.9\textwidth]{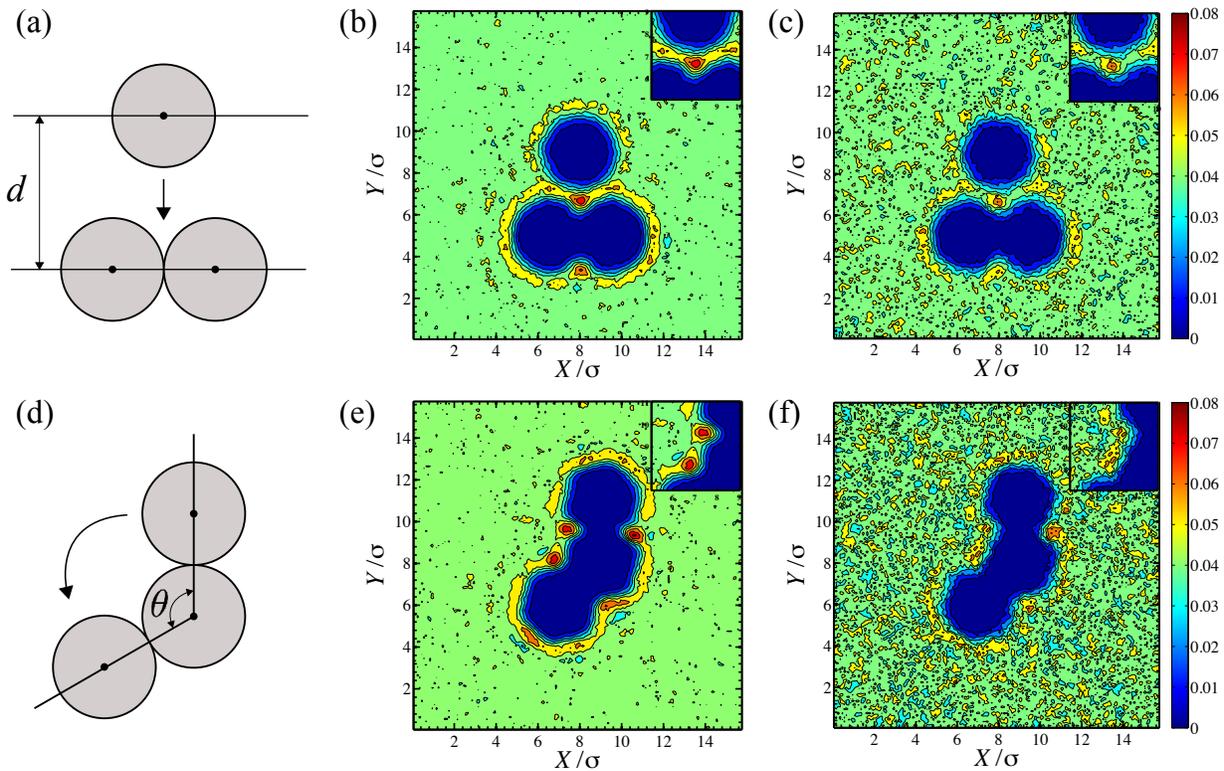}} 
}
\caption{(Color online) Understanding protein--protein interactions. (a) The third protein is placed at $d=4\,\sigma$ on the vertical symmetry line of the double-protein complex. (b) Density of lipid heads around RPs. (c) Density of lipid heads around SFPs. It is shown that lipids are more scattered and less concentrated at the triple contact region in (c) compared with (b). (d) The third protein is placed along an oblique line with angle $\theta=2\pi/3$ relative to the double-protein complex. (e) Density of lipid heads around RPs. The high concentration of lipids near the contact zones prohibits $\theta$ from tending to small angles if the angle constraint is released. (f) Density of lipid heads around SFPs. Lipids are scattered enough to allow for the rotation of $\theta$, and consequently, the formation of a triangular-shaped complex if the angle constraint is released. All insets: Closeups of contact regions.}
\label{fig6}
\end{figure*}

We explain the physical origin of different cluster-forming pathways of SFPs and RPs by investigating the distribution of lipids around proteins \cite{morozova2012shape, de2008molecular}. Our numerical simulations show that the lipid head and tail densities around a subject protein depend on the rigidity of the protein: RPs induce order and structure in surrounding lipids, which have formed ring-like structures around proteins (Fig. \ref{fig6}). The lipid density around rigid inclusions is generally lower around SFPs because the thermally vibrating flexible structure of relatively massive proteins continuously kicks and scatters lighter lipid molecules. For a complex of two proteins, the ordered structure of lipids takes an oval shape and constrains the formation of a complex with three proteins. 

To investigate how the process works, we designed two experimental scenarios. In the first scenario, two proteins are kept close to each other almost at the center of the lipid membrane by means of a spring of constant $500\,\epsilon/\sigma^2$. We then fix the position of the third protein on the double-complex vertical symmetry line at $d=4\,\sigma$ [Fig. \ref{fig6}(a)] with the same spring constant. The next step is to obtain the lipid distribution around the three-protein complex. When the triple-complex is composed of RPs, the high density of trapped lipids between the third protein and the double ones prohibits the formation of a triple, triangular-shaped complex as Fig. \ref{fig6}(b) shows. For SFPs, however, lipids are weakly bound to the double-complex, easily diffuse out of the triple-contact area, and thus facilitate the formation of bigger two-dimensional clusters if the constraint is released [Fig. \ref{fig6}(c)]. 

In the second scenario, the third protein is placed close to the double-complex and along an oblique line with an angle of $\theta=2\pi/3$ [Fig. \ref{fig6}(d)], forming a string-like structure. In the case of three RPs, lipids are trapped between two proteins and have completely filled the region between the two adjacent proteins. Therefore, these lipids resist the reduction of $\theta$ if the constraint is released. This is how RPs maintain their one-dimensional structure [Fig. \ref{fig6}(e)]. Note that in Figs. \ref{fig6}(a) and 6(d), the arrows show the most probable path that the third protein chooses to follow in order to form a triangular structure if the constraint is released. The distribution of lipids around SFPs is more homogeneous than that around RPs; compare Figs. \ref{fig6}(e) and 6(f). Our simulations with three proteins show that lipid molecules more easily diffuse in the space between SEPs and result in different arrangements of SEPs compared to RPs. More diffusive lipids around SFPs can also be identified by analyzing the mean squared displacement of lipids. For $c_p=0.48$, we have calculated the mean squared displacement for lipid molecules using the method in \cite{khoshnood2013anomalous}. Let us define $\rvec_{\parallel}(t)$ as the component of the position vector $\rvec(t)$ of lipid particles parallel to the bilayer surface. The two-dimensional diffusion coefficients $D=\lim_{t\rightarrow \infty}\langle \vert \rvec_{\parallel}(t)-\rvec_{\parallel}(0) \vert ^2\rangle /(4 t)$ that we find are approximately $43.1 \times 10^{-7}$ and $56.2\times 10^{-7} \, {\rm cm}^{2}/s$ for the ensemble of lipid molecules around RPs and SFPs, respectively. The diffusion coefficients are different from experimental data because the model is coarse grained. Here the relative change in diffusion coefficients is important. Our results show a $30\%$ reduction in diffusion coefficient for a membrane with RPs in comparison with one hosting SEPs. This is consistent with our observation of more restricted lipids surrounding RPs. We note that diffusion coefficients are measured from the part of the mean squared displacement profile that is linear in time. The initial anomalous region is not included in the calculations \cite{khoshnood2013anomalous}.     

\section{Discussion}
\label{sec:discuss}
Using coarse-grained molecular dynamic simulations, we showed that the rigidity of proteins has a profound effect on the cluster shape of TM proteins in lipid membranes. For proteins with zero hydrophobic mismatch, regardless of the protein concentration, RPs aggregate to form one-dimensional 
clusters while SFPs form two-dimensional clusters. In contrast to previous studies \cite{venturoli2005simulation, schmidt2008cluster, de2008molecular, west2009membrane} where hydrophobic mismatch is considered to be essential for clustering, we showed that proteins form stable clusters even in the absence of mismatch. Lipid-induced depletion interactions have been suggested as one of the main contributing factors to the interactions of cylindrical inclusions in bilayers \cite{west2009membrane,lague2001lipid,sintes1997protein}. This type of attraction occurs at distances smaller than the diameter of one lipid molecule. It has also been suggested that the interaction between membrane proteins largely depends on indirect lipid-mediated interactions \cite{west2009membrane}. Therefore, in this study, the interactions between proteins at short distances are most likely due to a strong depletion force of entropic origin explained by the Asakura-Oosawa model. This is a consequence of the high flexibility of the lipid chains. Direct protein--protein interactions can also play a role in the aggregation of proteins when protein particles are in the range of the cut-off distance. 

Khoshnood et al. \cite{khoshnood2010lipid} report on the same effect of depletion force for aggregation of rigid inclusions compared with completely flexible ones while hydrophobic mismatch exists. A key factor in association of proteins is the distribution of lipid molecules in close proximity to the inclusions \cite{yoo2013membrane}. These lipid molecules lose their entropy due to interaction with the lateral surface of proteins and form patterns around the inclusion. These patterns are more structured around RPs than SFPs; compare Figs. \ref{fig6}(b) and \ref{fig6}(e) with Figs. \ref{fig6}(c) and \ref{fig6}(f), respectively. The strong induced lipid structures around RPs lead to a low mobility of lipids, and consequently they have lower diffusion coefficients compared to lipids in a system with SFPs. As a result, the formation of a two-dimensional cluster, is obstructed by the induced lipid structure surrounding RPs, and in such systems one-dimensional aggregates are dominant.

We have observed fluctuation-induced attraction between two RPs and SFPs by applying the same constraint described in \cite{sintes1997protein}. This type of long-range interaction is caused by inclusions that affect the elastic properties of membranes and hence the fluctuation of lipid molecules. Long-range interactions play important roles in the clustering of both RPs and SFPs. However, the shape of the clusters is determined by a repulsive zone when proteins are close to each other and form groups of three or more proteins. The repulsive zone occurs when the distances between proteins are slightly larger than the size of one lipid molecule (beyond the depletion zone) \cite{lague2001lipid,sintes1997protein}. When a dimer forms, the fluctuation-induced attraction between proteins overcomes the effect of the repulsive zone and RPs and SFPs are able to create string-like aggregates at the early stages of simulations. The effect of the repulsive zone becomes more prominent when RPs want to create two-dimensional aggregates and can not be overcome by the fluctuation-induced attraction between inclusions. In the case of SFPs, the fluctuation-induced attraction overcomes the effect of the repulsive zone and pushes the third protein into the depletion zone. Both the fluctuation-induced attraction and the repulsive zone are classified as lipid-mediated interactions.

Cell adhesion to extracellular matrices is regulated by the size and position of focal adhesions and, most importantly, how they are distributed \cite{elineni2011regulation}. The latter is controlled by integrin protein association. Integrin has both $\beta$ and $\alpha$ subunits and the number of $\beta$ and $\alpha$ subunits varies depending on the type of integrin, which means a variety of structural rigidity. According to our results variation in integrin stiffness affects their aggregation patterns and consequently may have an impact on the quality of cell adhesion. 

Cell receptor activation depends on receptors' clustering and their conformational changes. The latter requires rearrangements of proteins in the cluster \cite{minguet2007full}. It means that the pattern in which receptors are attached to each other in a cluster plays a role in the activation process. Receptors may have subunits with different structural rigidities, and based on our findings this feature can affect their final aggregation pattern. 
 
Remodeling of the biomembrane is achievable by proteins \cite{simunovic2013linear} and nanoparticles \cite {vsaric2012fluid}, and it is a vital step in endocytosis and vesiculation \cite {reynwar2007aggregation}. Interestingly, in a system of spherical nano-particles on lipid membranes \cite {vsaric2012fluid} different aggregation patterns have been induced by varying the membrane rigidity. The majority of living cell membranes are made up of phospholipids and the rigidity of the bilayer is known. Our results suggest that flexibility of the inclusion may work as a controlling parameter for membrane remodeling when we can not alter the membrane rigidity. The new controlling parameter can also help in the design of therapeutic peptides to tackle protein-aggregation diseases. This is obtainable by protein engineering methods without perturbing significantly the overall stability or activity of the protein \cite{villegas2000protein}. Different mixtures of amino acids have different amounts of alpha-helical and beta-structural units and consequently 
have different rigidity \cite{perticaroli2014rigidity}. Therefore, it is possible to design proteins with a specifically defined rigidity and certain aggregation behavior, which lead them to perform a specific biological function. 

The protein model in our study is a toy model with SFPs and RPs resembling $\alpha$-helices and $\beta$-sheets, respectively, as their structural rigidities differ substantially \cite{perticaroli2014rigidity}. The simulations by Parton et al. \cite{parton2011aggregation} with $\alpha$-helical and $\beta$-barrel proteins on vesicles and flat membranes show that while $\alpha$-helical proteins form two-dimensional clusters, $\beta$-barrel proteins constitute linear aggregates. In their experiments, a combination of several factors such as hydrophobic mismatch, membrane curvature, and the shape and class of proteins can be held responsible for the observed discrepancies in the aggregation patterns of proteins. Our results demonstrate that structural rigidity as a sole factor determines the shape and pattern of protein aggregates.

\bibliography{ref}

\end{document}